\documentclass[12pt]{article}
\usepackage{graphicx}

\newcommand\pubnumber{WSU--HEP--XXYY}
\newcommand\pubdate{\today}

\def\babar{\mbox{\sl B\hspace{-0.4em} {\small\sl A}\hspace{-0.37em} \sl B\hspace{-0.4em} {\small\sl A\hspace{-0.02em}R}} }
\def\ific{IFIC, Instituto de F\'{i}sica Corpuscular (CSIC - U. Valencia), Parque Cient\'{i}fico, C/Catedr\'{a}tico Jos\'{e} Beltr\'{a}n, 2,  E-46980 Paterna, Spain}

\def\pienu{$D^0 \to \pi^- e^+ \nu$ }
\def\ff{$f_{+,D}^{\pi}(q^2)$ }
\def\Title#1{\begin{center} {\Large #1 } \end{center}}
\def\Author#1{\begin{center}{ \sc #1} \end{center}}
\def\Address#1{\begin{center}{ \it #1} \end{center}}

\newcommand\pubblock{\rightline{\begin{tabular}{l} \pubnumber\\
         \pubdate  \end{tabular}}}
\newenvironment{Abstract}{\begin{quotation}  }{\end{quotation}}
\newenvironment{Presented}{\begin{quotation} \begin{center} 
             PRESENTED AT\end{center}\bigskip 
      \begin{center}\begin{large}}{\end{large}\end{center} \end{quotation}}

\def\beq{\begin{equation}}
\def\eeq#1{\label{#1}\end{equation}}
\def\eeqn{\end{equation}}

\def\beqa{\begin{eqnarray}}
\def\eeqa#1{\label{#1}\end{eqnarray}}
\def\eeqan{\end{eqnarray}}

\let\bar=\overbar

\def\Dslash{\not{\hbox{\kern-4pt $D$}}}
\def\dslash{\not{\hbox{\kern-2pt $\del$}}}

\def\msb{{\bar{\ssstyle M \kern -1pt S}}}

\begin{document}
\begin{titlepage}
\pubblock

\vfill
\Title{Measurement of the $D^0 \to \pi^- e^+ \nu$ branching fraction, form factor and implications for $V_{ub}$}
\vfill
\Author{Arantza Oyanguren}
\Address{\ific}
\vfill
\begin{Abstract}
In this talk results of the study of the $D^0 \to \pi^- e^+ \nu$ decay channel, recorded by the \babar detector at the c.m. energy close to 10.6 GeV, are reported. The branching fraction of this channel is measured relative to the $D^0 \to K^- \pi^+$ decay. The hadronic form factor,  $f_{+,D}^{\pi}(q^2)$, function of $q^2$, the four momentum transfer squared between the $D$ and the $\pi$ mesons, is compared to various theoretical predictions, and the normalization $V_{cd} \times f_{+,D}^{\pi}(q^2=0)$ is extracted from a fit to data. Results are compared with Lattice QCD calculations. A new multipole model is applied which makes use of present information of resonant states contributing to the form factor. With the understanding of the $f_{+,D}^{\pi}(q^2)$ form factor, and provided the relation between the $D^0 \to \pi^- e^+ \nu$  and $B^0 \to \pi^- e^+ \nu$ decay widths at the same pion energy, the CKM matrix element $V_{ub}$ is determined and compared to recent measurements. This method of extracting $V_{ub}$ will become competitive with new Lattice QCD calculations of the ratio of form factors.   
\end{Abstract}
\vfill
\begin{Presented}
The 7th International Workshop on Charm Physics (CHARM 2015)\\
Detroit, MI, 18-22 May, 2015
\end{Presented}
\vfill
\end{titlepage}
\def\thefootnote{\fnsymbol{footnote}}
\setcounter{footnote}{0}
%

\section{Motivation}
The differential decay width of the \pienu decay channel\footnote{Charge conjugated are implicit in this document} as function of $q^2$, the four momentum transfer squared between the $D$ and the $\pi$ mesons, can be expressed in terms of the Cabibbo-Kobayashi-Maskawa (CKM) matrix element, $V_{cd}$, and, neglecting the electron mass, a unique form factor, \ff. This form factor describes all the non-perturbative QCD effects in the $D \to \pi$ transition:
\begin{equation}
 \frac{d \Gamma(D^0 \to \pi^- e^+ \nu)}{d q^2} = \frac{G^2_F}{24 \pi^3}\left ( V_{cd} 
\times |f_{+,D}^{\pi}(q^2)| \right )^2 p^{*3}_\pi(q^2),
\label{eq:diff_decay_rate}
\end{equation}
$p^{*}_\pi$ being the pion momentum in the $D^0$ rest frame. 
 The form factor can be represented as an infinitive sum of poles, corresponding to resonant states, $D_i^{*}$, with $J^P = 1^-$ which couple to $D\pi$. 
\begin{eqnarray}
f_{+,D}^{\pi}(q^2) \simeq 
\sum_{n=0}^{\infty}\frac{Res(f_{+,D}^{\pi})_{D_i^{*}}}{m^2_{D_i^{*}}-q^2}.
\label{eq:dispers2}
\end{eqnarray}
For the \pienu decay channel one has the advantage that part of \ff is known, since contributions from the leading state $D^\ast$ and the first radially excited state $D^{\ast '}$ are known and can be used to constrain the form factor \cite{ref:pienu,ref:damirfdstar}. Another interest comes from the fact that the \pienu and  $B^0\to \pi^- e^+ \nu$ decay channel can be related at the same pion energy, allowing, if the form factors are known, the extraction of the CKM matrix element $V_{ub}$.   
\section{Analysis method}
\begin{figure}[!htb]
  \centering
\includegraphics[height=7.6cm]{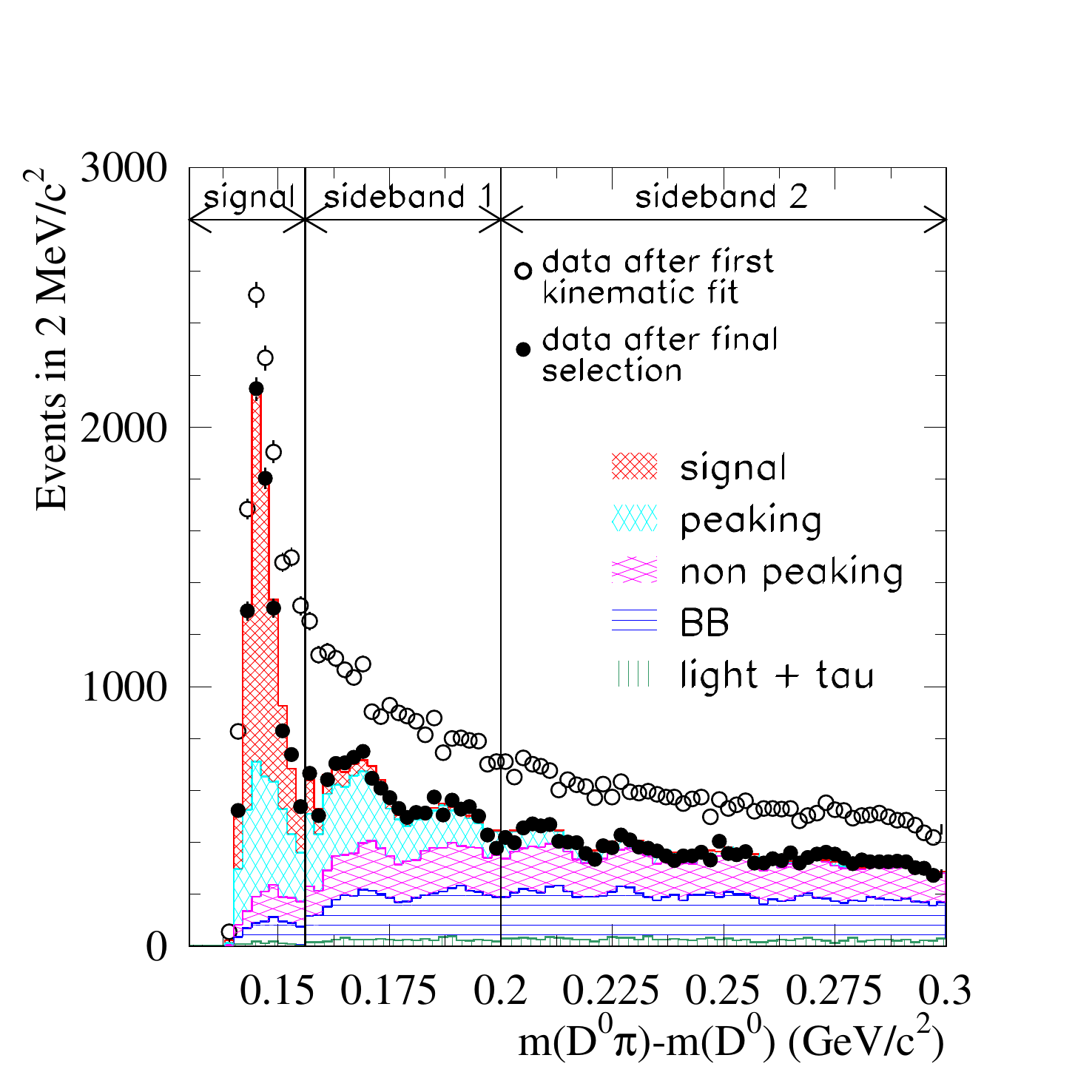}
\includegraphics[height=7.4cm]{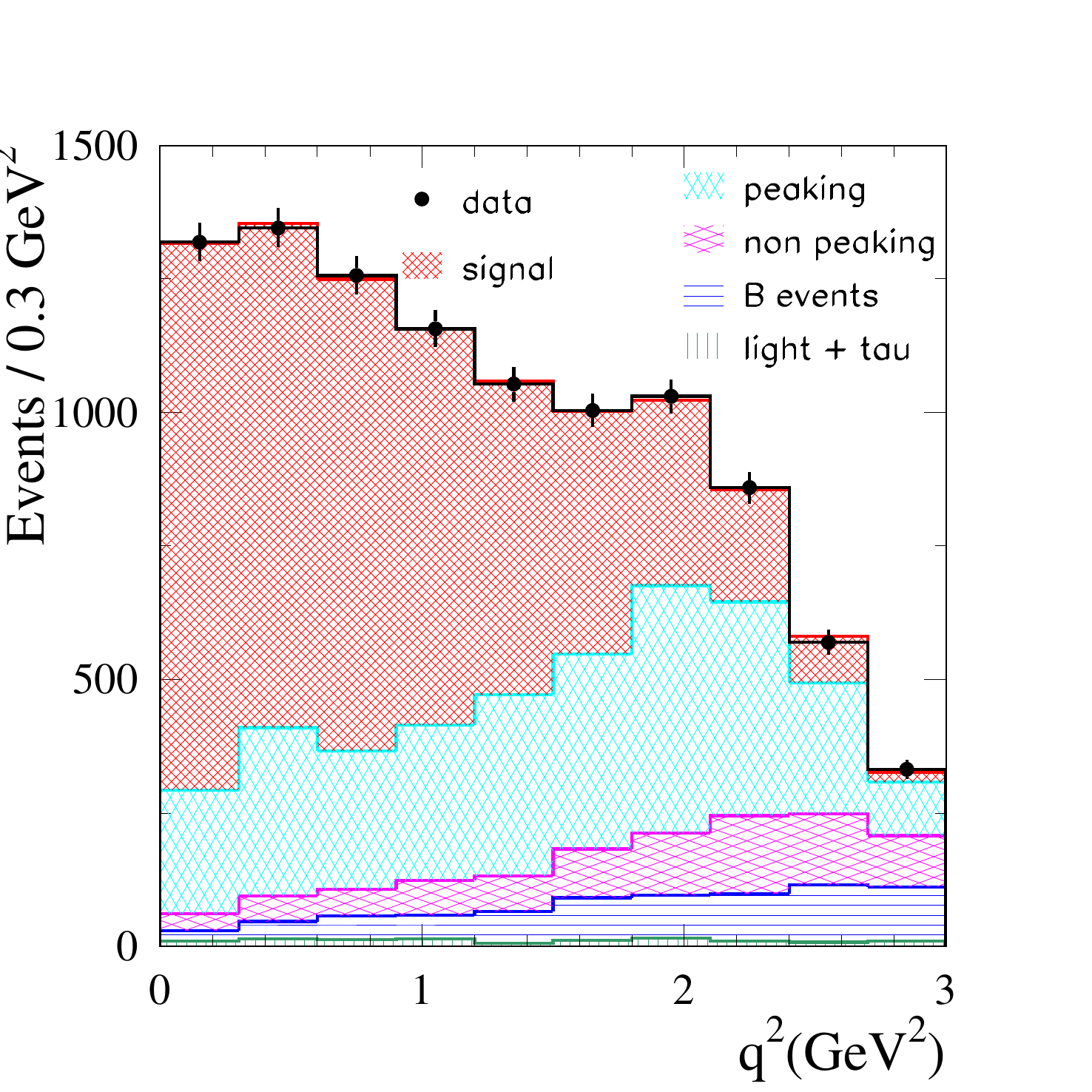}
  \caption[]{ Left: mass difference $\delta(m) = m(D^0 \pi^+)-m(D^0)$. Open and full circles correspond to kinematic fits applying the $D^0$ and $D^*$ constraints \cite{ref:pienu}. The distribution for MC-simulated signal and the different background distributions are superimposed. Right: the measured $q^2$ distribution (data points) 
for events selected in the $\delta(m)$ signal region is compared to the sum of 
the estimated backgrounds and the fitted signal component.}
\label{fig:deltam}
\end{figure}

This analysis \cite{ref:pienu} is based on similar techniques as other charm semileptonic decays at \babar \cite{ref:kenu,ref:dskkenu,ref:dkpienu}. The \pienu decay channel has the difficulty that it is Cabibbo-suppressed, with a small branching fraction, and suffers in addition from large background from pions. Using 347.2 fb$^{-1}$ of $e^+e^- \to c\bar c$ data recorded by the \babar detector at the $\Upsilon(4S)$ energy, the decay $D^{*+} \to D^0 \pi^+$ with \pienu is reconstructed using a partial reconstruction technique. The two pions and the positron are reconstructed in the same event hemisphere, requiring a tight particle identification for signal pions and vetoing kaons. The $D^0$ four momentum is obtained from the reconstructed particles and the missing energy from the information of the rest of the event. Constraints on the $D$ and $D^*$ masses are applied in a kinematic fit to obtain the $q^2$ distribution. This method is validated using hadronic $D^0 \to K^-\pi^+$  data. The main issue of the analysis concerns the background suppression. Using Fisher discriminant variables against $B\bar B$ and charm backgrounds, the S/B rate is about 1.2, with a signal efficiency around 1.8$\%$.
To further control the several sources of background events, the mass difference between the $D^*$ and the $D^0$ is used.
The signal region is defined as $\delta(m) < 0.155~{\rm GeV}^2$. Two additional  $\delta(m)$  windows are used to evaluate the background contributions. Using the event missing energy and the pion momentum information, the rates for the different types of background events are constrained. In this way, the main source of systematic uncertainty in the analysis, due to the background control, is assessed using data. The $\delta(m)$ distribution is shown in Fig.~\ref{fig:deltam}-left. 
The $q^2$ distribution, $q^2=(p_{D^0}-p_{\pi^{-}})^2 = (p_{e^+}+p_\nu)^2$ is measured then in 10 bins and a fit to data is applied using several parameterizations of the form factor. In Fig. \ref{fig:deltam}-right the $q^2$ data distribution is shown. The MC-simulated events, with corrected background components is also shown. It is verified that the angular distribution is well reproduced for the signal and background components. 

\pienu events are normalized to the number of $D^0 \to K^- \pi^+$ decay events, trying to have a selection as similar as possible for the semileptonic and hadronic channels. The ratio $R_D = \frac{ {\cal B} (D^0 \to \pi^- e^+ \nu)}{ {\cal B} (D^0 \to K^- \pi^+)}$ is measured to be $R_D = 0.0702 \pm 0.0017 \pm 0.0023$, where the first uncertainty is statistical and the second one is systematic. 
Using the world average for the $ {\cal B} (D^0 \to K^- \pi^+)$ \cite{ref:hfag12}, the branching fraction of the \pienu decay channel is measured to be ${\cal B}(D^0 \to \pi^- e^+ \nu) = (2.770 \pm 0.068 \pm 0.092 \pm 0.037) \times 10^{-3}$, where the third uncertainty comes from the uncertainty in the normalization channel. 

\section{Form factor interpretation}
\begin{figure}[!htb]
  \centering
\includegraphics[height=6.5 cm]{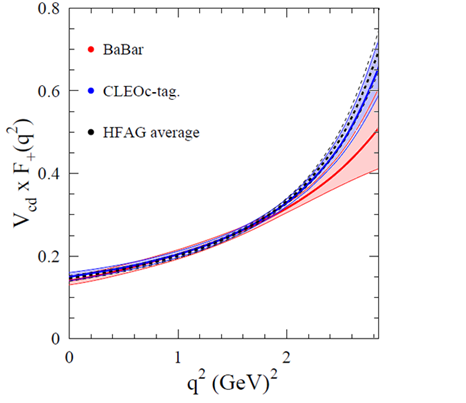}
\includegraphics[height=6.5 cm]{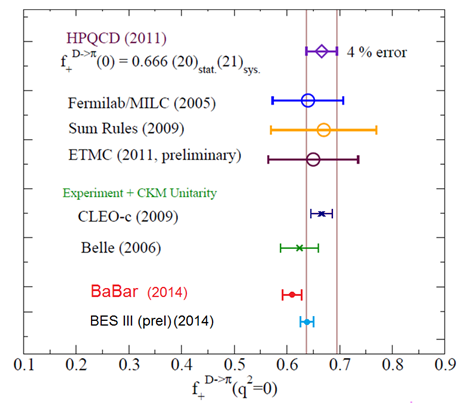}
  \caption[]{Left: form factor fit using the $z$-expansion for \babar \cite{ref:pienu}, CLEO-c \cite{ref:cleoc09} and the HFAG average \cite{ref:hfag12}. Bands represent the fit uncertainty. Right: Comparison of  $f_{+,D}^{\pi}(0)$ obtained by different experiments and by Lattice-QCD calculations.}
\label{fig:comparison}
\end{figure}

Having measured the number of \pienu events as function of $q^2$ (Fig. \ref{fig:deltam}-right), several parameterizations can be tried to describe and fit the form factor. One of the most extensively used is called the $z$-expansion,
a model-independent parametrization based on general properties of QCD~\cite{ref:beforehill}.  
In terms of the variable $z$, defined as 
\begin{equation}
z(t,t_0) = \frac{\sqrt{t_+-t} - \sqrt{t_+-t_0}}{\sqrt{t_+-t} + \sqrt{t_+-t_0}} ,
\end{equation}
where $t\equiv q^2$, $t_0=t_+(1 - \sqrt{1-t_-/t_+})$ and $t_\pm=(m_{D^0}\pm m_{\pi^+})^2$, 
the form factor,  takes the form:
\begin{equation}
f_{+,D}^{\pi}(t)=\frac{1}{P(t) \Phi(t,t_0)}\sum_{k=0}^{\infty}a_k(t_0)~ z^k(t,t_0) ,
\label{eq:taylor}
\end{equation}
where $P(t)=1$ and $\Phi(t,t_0)$ is an arbitrary analytical function. A standar choice for $\Phi(t,t_0)$ is used by the different experiments \cite{ref:pienu}. The fitted parameters are commonly defined as $r_k = a_k / a_0$ for $k = 1,2$,  and the overall normalization of the expansion is $V_{cd} \times f_{+,D}^{\pi}(0)$.
Results of the fit to \babar data are: $r_1 = -1.31 \pm 0.70 \pm 0.43$, $r_2 = -4.2 \pm 4.0 \pm 1.9$, and $V_{cd} \times f_{+,D}^{\pi}(0) = 0.1374 \pm 0.0038 \pm 0.0022 \pm 0.0009$. The last uncertainty in the normalization comes from external inputs \cite{ref:pienu}. 
The main disadvantage of this parametrization is that the $a_k$ parameters have no physical meaning and cannot provide an interpretation of the form factor. It is difficult to constrain the contribution from the $D^{*+}$ pole in this parameterization because it requires extrapolation beyond the physical region. Fig. \ref{fig:comparison} compares the \babar result with several measurements at different experiments and with Lattice QCD calculations. Assuming unitarity of the CKM matrix, with $|V_{cd}| = |V_{us}| = 0.2252 \pm 0.0009$ \cite{ref:pdg2013}, the \babar result leads to 
$f_{+,D}^{\pi}(0) = 0.610 \pm 0.017 \pm 0.010 \pm 0.005$. If, instead, one uses the Lattice QCD results on the form factor,  
$f_{+,D}^{\pi}(0) = 0.666 \pm 0.029$ \cite{ref:flag}, the $V_{cd}$ matrix element results in $V_{cd} = 0.206 \pm 0.007 \pm 0.009$. 

\begin{figure}[!htb]
  \centering
\includegraphics[height=6.5 cm]{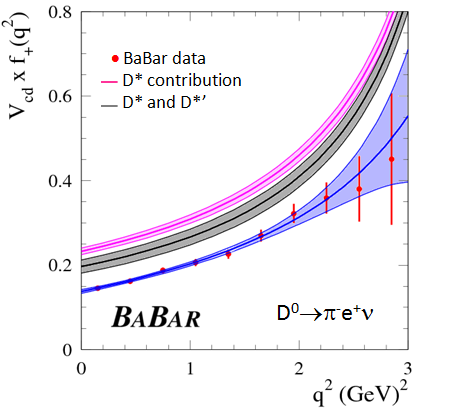}
\includegraphics[height=7.3 cm]{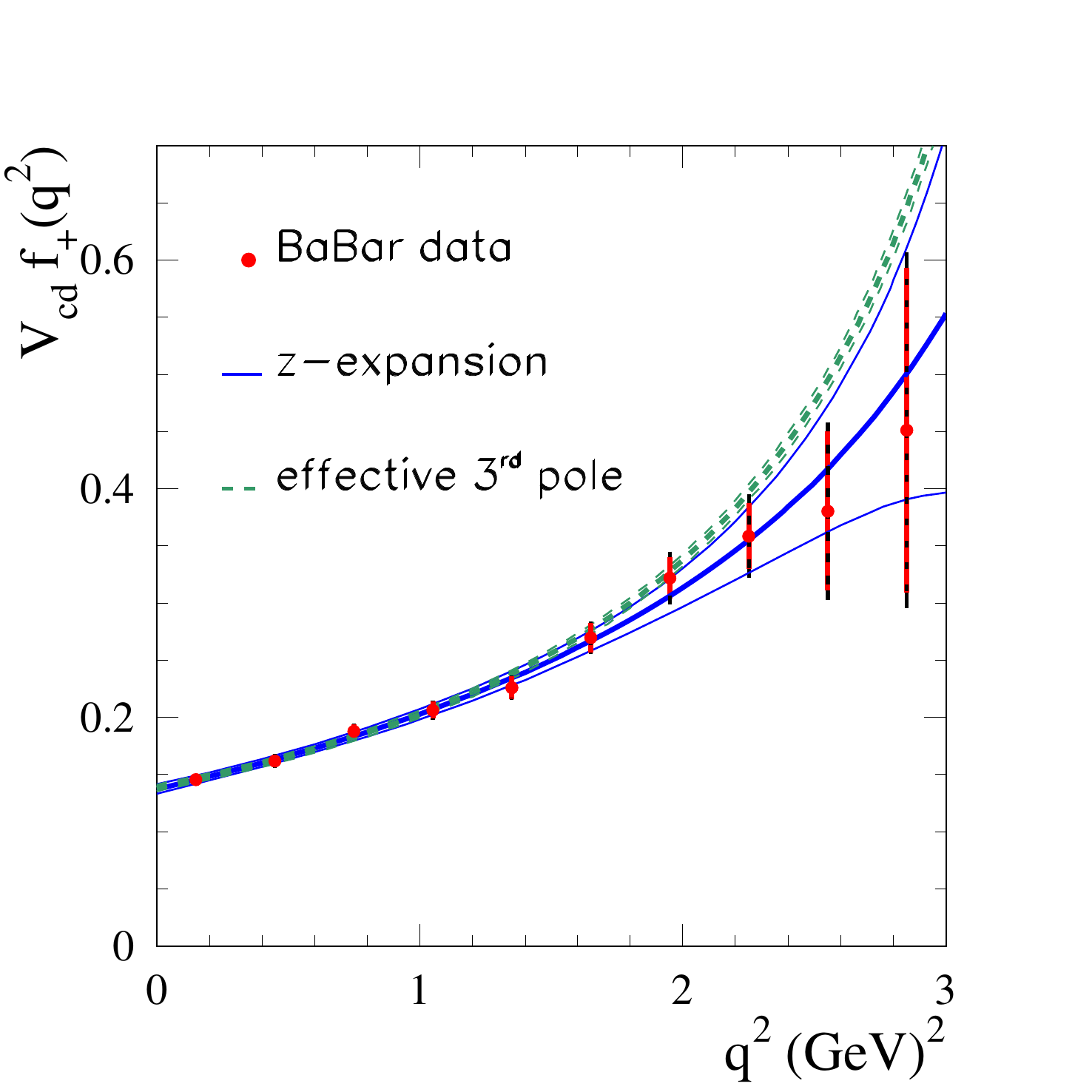}
  \caption[]{Left: Measured $f_{+,D}^{\pi}(q^2)$ at \babar and expected contributions of the $D^*$ and $D^{*'}$ poles to the form factor. Bands represent the uncertainties. The gap between data points and the gray curve reveals the need of additional hadronic states to describe the form factor. Right: \babar data and fitting curves using the z-expansion (solid blue) and the {\it three-pole} ansatz (dashed green).}
\label{fig:pole_contributions}
\end{figure}

However, one can go further in the understanding of the \pienu form factor: based on \cite{ref:burdman} and \cite{ref:bk}, a {\it three-poles} ansatz has been developed \cite{ref:damirfdstar}. The form factor can be expressed as an infinitive sum of $J^P=1^-$ states (Eq. \ref{eq:dispers2}). The residue which defines the contribution of these states ($D^{*}_{i}$ resonances) can be, in turn, expressed in terms of the meson decay constants $f_{D^{*}_{i}}$ and coupling to the $D \pi$ final state, $g_{D^{*}_{i}D\pi}$: 
\begin{equation}
Res(f_{+,D}^{\pi})_{D^{*}_{i}}=\frac{1}{2} m_{D^{*}_{i}} f_{D^{*}_{i}}g_{D\pi} . 
\label{eq:residue}
\end{equation}
For the \pienu decay channel, the decay constants for the two first states, the leading $D^*$ meson and the first radially excited state, $D^{*'}$, have been computed by Lattice QCD \cite{ref:damirfdstar}. The $g_{D^{*(')} D\pi}$ couplings can be obtained from the experimental results at \babar of the measured widths for these resonances \cite{ref:babardstar,ref:dstarprime}. In this way the contribution of these states to the form factor is well defined and constrains $f_{+,D}^{\pi}(q^2)$. In Figure \ref{fig:pole_contributions}-left the contributions of the $D^*$ and the $D^{*'}$ are given, together with the form factor fit to \babar data using the $z$-expansion formalism. This plot reveals the fact that the form factor cannot be explained by only these two contributions, and then additional poles are needed to fill the gap between the measured data and the $D^*+D^{*'}$ contribution (gray curve). Using the constraints given by the  $D^*$ and the $D^{*'}$ poles, one can consider an effective third pole contributing to the form factor \cite{ref:damirfdstar}. The {\it superconvergence} condition, $\sum_{i} Res(f_{+,D}^{\pi})_{D^{*}_{i}} \simeq 0$ , is applied, from the behaviour of the form factor at very large values of $q^2$ \cite{ref:burdman}. 
In Figure \ref{fig:pole_contributions}-right, the fit to \babar data using this description of the form factor is presented. Once should note that data is well described by this ansatz. The effective third pole mass is fitted and results in the value $m_{D^{*''}_{eff}} = (3.6 \pm 0.3)$~GeV , which is larger than the predicted third $J^P=1^-$ state by quarks models ($\sim 3$~GeV), as it is expected since it is considered as an effective pole. A unique contribution from this predicted state is excluded by data. 

\section{$V_{ub}$ extraction}
Having measured $d\Gamma_{D^0\to \pi^-e^+\nu}/dq^2$, one can extract the CKM matrix element $V_{ub}$ from the relation between the 
\pienu and $B^0\to\pi^-e^+\nu$ decay channels, valid for a common range in the energy of the ejected pion in the rest frame of the heavy-light meson, which is less than about $E^*_\pi \sim 1$~GeV.  Instead of $E^*_{\pi}$, one can use the Lorentz invariant variable $w_{D(B)} = v_{D(B)} \cdot v_{\pi}$, where $v_{D(B)}=p_{D(B)}/m_{D(B)}$ and $v_{\pi}=p_{\pi}/m_{\pi}$ 
are the four-velocities of the $D(B)$ and $\pi$ mesons, respectively. In terms of this quantity: $q^2 =  m_{D(B)}^2 + m_{\pi}^2 -2~m_{D(B)}m_{\pi}w_{D(B)}$, and then at $w_D = w_B$:
\begin{equation}
\frac{d \Gamma^B /d w_B }{d \Gamma^D /d w_D }= \frac{m_B}{m_D} \left (\frac{V_{ub}}{V_{cd}} \right )^2 \left |\frac{f_{+,B}^{\pi}(w_B)}{f_{+,D}^{\pi}(w_D)}\right |^2.
\label{eq:ratio_width}
\end{equation}
The $V_{ub}$ element can be obtained from Eq.(\ref{eq:ratio_width}) if the ratio between the $f_{+,B}^{\pi}(w_B)$ and $f_{+,D}^{\pi}(w_D)$ form factors is known. This ratio can be obtained from Lattice QCD calculations, or using a phenomenological model. 
The experimental common range for the two decays on $w_{B, D}$ is between 1 and 6.7. This corresponds to $q^2$ from 0 to 2.975~GeV$^2$ for $D$ decays and from 18 to 26.4~GeV$^2$ for B decays, leading only about $17\%$ of overlapping region. Nevertheless, a physics interpretation of the charm form factor allows to use it outside the $D$ physical region and to determine the form factor ratio.   

\begin{figure}[!htb]
  \centering
\includegraphics[height=6.8cm]{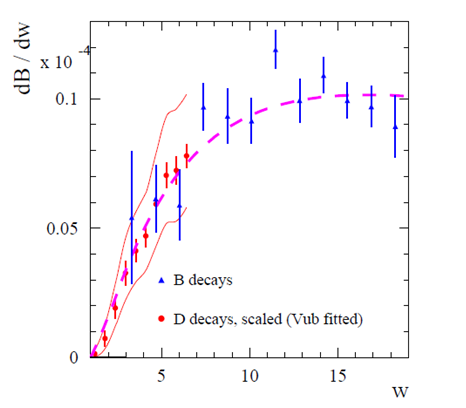}
\includegraphics[height=7.3cm]{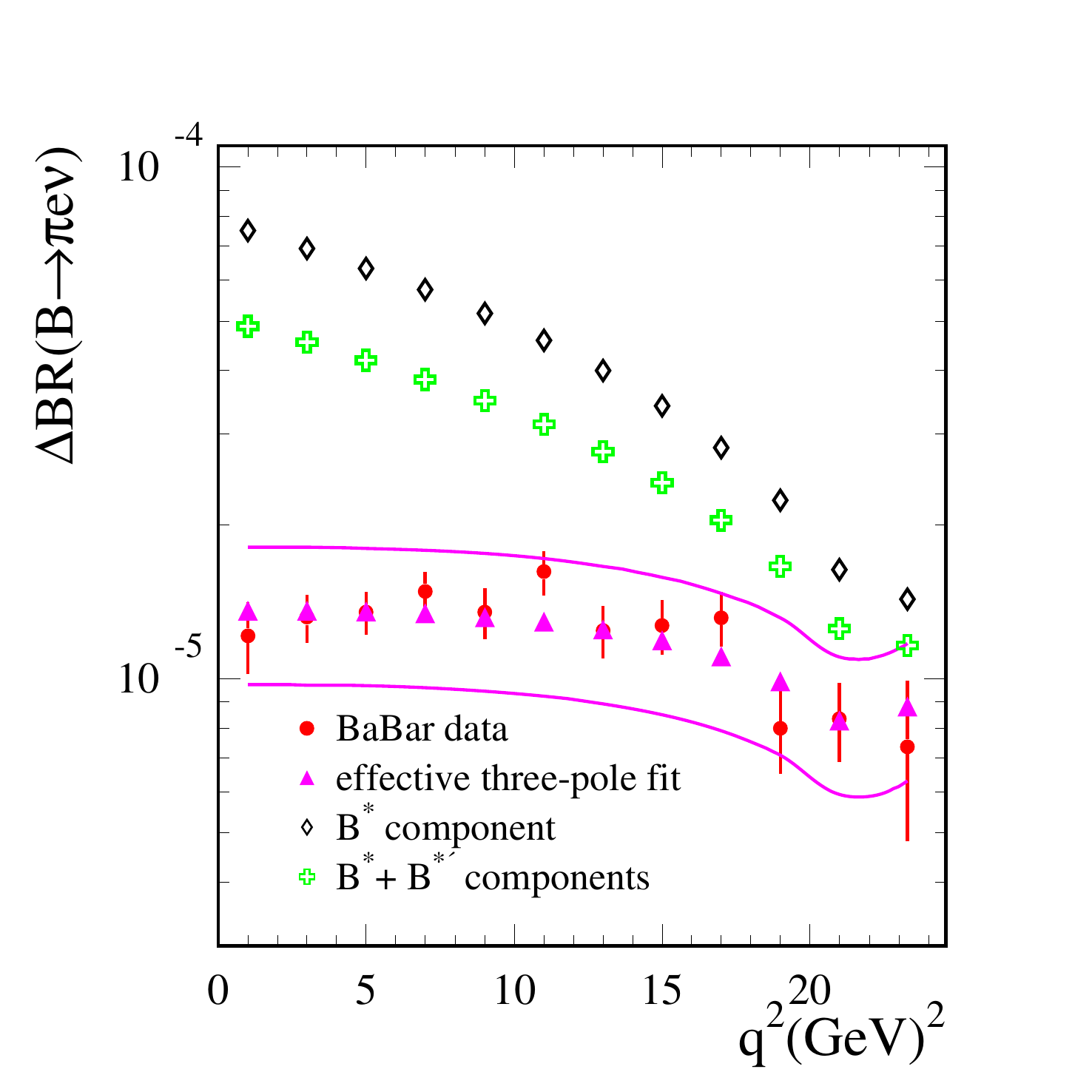}
  \caption[]{ Left: Fit to the $d{\cal B}/dw$ differential decay rate for $B^0\to\pi^-e^+\nu$ events measured by \babar with an extrapolation of the \pienu form factor measurement, assuming a constant ratio of form factors equals to $1.8\pm0.2$ from Lattice calculations. The $V_{ub}$ parameter is fitted. Right: Fit to the $B^0\to\pi^-e^+\nu$ events using the {\it three-poles} phenomenological model for the form factor, using constraints for the residues of $B^*, B^{*'}$ and fitting a third pole with an effective mass. The contribution for each pole is shown. The two lines  indicate the theoretical uncertainties.}
\label{fig:vub}
\end{figure}

Two different approaches have been used to extract $V_{ub}$, leading to systematic uncertainties of very different origin. \babar data for \pienu  \cite{ref:pienu} and $B^0\to\pi^-e^+\nu$ decays \cite{ref:babarb} are used:

- $V_{ub}$ from Lattice results of individual form factors:\\
Considering the Lattice results for $f_{+,B}^{\pi}(q^2)$ \cite{ref:hpqcd3,ref:milcfnal1}
and  $f_{+,D}^{\pi}(q^2)$ \cite{ref:hpqcd2} one observes that the two form factors have a similar $w$ dependence. For $w>4$, the ratio of these form factors is $1.8\pm0.2$\footnote{One should note that this value is consistent with the expectation at first order: $\sqrt{m_B/m_D}=1.7$.}. Assuming this constant value, and extrapolating to the nonphysical region the measured \pienu form factor, in terms of the {\it three-poles} model, the $V_{ub}$ matrix element is fitted, as it is shown in Fig. \ref{fig:vub}-left, giving the value:
 \begin{center} $V_{ub} = (3.65 \pm 0.18 \pm  0.40)\times 10^{-3}$.\end{center}
The dominant contribution to the systematic uncertainty originates from the form factor ratio from Lattice. This result can be improved if Lattice QCD provides values for this ratio with better accuracy and for several values of $q^2$. 

- $V_{ub}$ from the {\it three-poles} phenomenological model:\\
Another approach consists in fitting directly $B^0\to\pi^-e^+\nu$ events using the {\it three-pole} phenomenological model for the $B\to \pi$ form factor, since it has been proven to work well for the \pienu decay. For that one can use constraints from the residues for the first two poles, which correspond to the $B^*$ and $B^{*'}_{1}$, taking the mass of the latter from \cite{ref:defazio}, and fitting the third pole with an effective mass. The result of this third effective pole is $m_{B^{*'}_{eff}} = (7.4 \pm 0.4)$~GeV. It is expected that the ratio for the residue at the different poles are the same for $D$ and $B$ semileptonic decays \cite{ref:burdman}, and a constraint is applied in the fit. The fit results in:  
  \begin{center} $V_{ub} = (2.6 \pm 0.2 \pm  0.4)\times 10^{-3}$,\end{center}
and it is shown in Figure \ref{fig:vub}-right. The uncertainty on $V_{ub}$ is dominated in this case by the knowledge of the $g_{B^{*(')}B\pi}$ couplings entering in the residue. This value could also be improved by new Lattice calculations.  
 
\section{Conclusions}
The \pienu form factor and branching fraction is measured at \babar. Results are competitive and in agreement with CLEO-c, BELLE and preliminary results from BES III. Experimental results in this channel are at present more accurate than Lattice QCD calculations.  
 A physics interpretation of the $D\to \pi$ form factor is developed, in terms of a phenomenological {\it three-poles} model, using precise information of the two first contributing poles. It is observed that these two poles cannot explain alone the form factor and an effective third pole is obtained. This description agrees well with data. 
The $V_{ub}$  matrix element is extracted using charm semileptonic decays, through two alternative approaches: assuming a constant form factor ratio from Lattice QCD, or considering the {\it three-poles} model, proved on \pienu decays,  on  $B^0\to\pi^-e^+\nu$ events. Results are in agreement with recent Lattice calculations \cite{ref:lattice_newvub} and the measurement from LHCb \cite{ref:vub_lhcb}, in particular for the first approach. This method of $V_{ub}$ extraction will become competitive if new Lattice QCD computations on the form factor ratio are available.

\end{document}